# Sensitivity of health-related scales is a non-decreasing function of their classes.

Vasileios Maroulas[1] and Demosthenes B. Panagiotakos[2,*]

[1] Institute for Mathematics and its Applications, University of Minnesota, Minneapolis, USA. Email: maroulas@ima.umn.edu

[2] Office of Biostatistics & Epidemiology, Department of Nutrition Science - Dietetics, Harokopio University, 70 Eleftheriou Venizelou Str., 176 71, Athens, Greece. Email: dbpanag@hua.gr Tel., +30 210 9549 332 & Fax., + 30 210 9600 719.

*corresponding author

# Abstract.

In biomedical research the use of discrete scales which describe characteristics of individuals are widely applied for the evaluation of clinical conditions. However, the number of classes (partitions) used in a discrete scale has never been mathematically evaluated against the accuracy of a scale to predict the true cases. This work, using as accuracy markers the sensitivity and specificity, revealed that the number of classes of a discrete scale affects its estimating ability of correctly classifying the true diseased. In particular, it was proved that the sensitivity of scales is a non-decreasing function of the number of their classes. This result has particular interest in clinical research providing a methodology for developing more accurate tools for disease diagnosis.

# 1. Introduction

In biomedical literature, discrete variables, usually named scales or indices, are widely used for describing individuals' characteristics, e.g. the level of depressive and anxiety symptoms, the quality of diet, the sense of pain, for the evaluation of various clinical conditions or behaviors, like obesity, hypertension, diabetes, cardiovascular disease risk (Kant 1996, Kourlaba and Panagiotakos 2009). Despite the fact that health-related scales are an important tool for the assessment of the relationships between individuals' characteristics and their health status, the methodology of their construction is not utterly clarified. For example, it is unknown which is the optimal number of values of the scale (i.e., binary, small range, such as <5 classes or large range), if continuous ranking (i.e., 1, 2, 3, 4, …) or discontinuous (i.e., 1, 2, 5, 6, 7, 10, …) is better, whether special weights are needed for each class and other issues that may improve the diagnostic accuracy of the scale in evaluating a health outcome. At this point it should be noted that categorization of a continuous variable results in a loss of information and consequently, reduction in the ability of the independent variable to accurately predict the dependent. At first, the latter has not analytically been evaluated in terms of diagnostic ability of a scale; while, secondly, in real practice use of continuous scales to measure individual's behavioural attributes is difficult to apply in practice (e.g., it is difficult or even impossible to accurately measure emotions, sense of pain, etc using continuous scales, because of the increased likelihood of misclassification). Thus, the use of discrete, and particularly small-range (because of potential misclassification) scales is usually the best approach used in practice.

Recently, and based on both simulated and empirical data it has been shown that increasing the number of classes, the sensitivity (*Se*) of the scale is increased, too (Kourlaba and Panagiotakos 2009b). Specifically, using composite scales that are summations of discrete random variables it was revealed that the diagnostic accuracy of the scale increases as the number of classes of scale components increases, as well. Moreover, it was observed that the use of continuous components is the optimal for achieving the maximum diagnostic ability. Therefore, the findings of this work are of major importance for risk assessment tools in related research since they strongly suggest using as many categories as it can be used for a scale component. However, to the best of our knowledge, a mathematical proof of the aforementioned hypothesis has never been presented.

### 1.1 Aim

The hypothesis tested in this work was whether the number of classes of an ordinal, discrete variable, $X$, affects its ability for better predicting a response binary variable $Y$ that plays the role of a response outcome. The predicting ability was evaluated through the sensitivity, $Se$, of the scale.

The strategy followed was by considering two random discrete variables $X^k$ and $X^{k+1}$ with $k$ and $k+1$ classes, respectively; and one random variable $Y$ from the Bernoulli distribution. Defining $(Se)^k = P[X^k > c \mid Y=1]$ and $(Se)^{k+1} = P[X^k > c' \mid Y=1]$ the sensitivity of the $X^k$ and $X^{k+1}$ variables, respectively ($c$ and $c'$ are the thresholds where the probability of having $Y=1$ is maximized), we prove that $(Se)^k \leq (Se)^{k+1}$.

## 2. Proof

This section provides a proof of our statement. A mathematical definition of the sensitivity is given below.

***Definition 1:*** *Let Y be a random variable distributed via Bernoulli(p). For a random variable X, positively associated with $p = P[Y=1]$, the sensitivity of X, Se(X), in relation to Y is defined as the conditional probability,*

$$Se(X) = P[X > c \mid Y=1], \quad (1)$$

where c is some constant.

**Remark 1:** The specificity of X, Sp(X), in relation to Y, is another criterion for measuring the predicting ability of a model. The specificity is defined as the conditional probability, $Sp(X) = P[X<c \mid Y=0]$. The c value, used in (1), is a threshold that belongs to the domain of X. In practice, c is the value of X that maximizes the joint probabilities $\{Se(X), 1-Sp(X)\}$.

Let **(Ω, P)** be a probability space, and $X^k: \Omega \to D^k$ be a discrete random variable, where $D^k = \{x_1, x_2, ..., x_k\}$. Given Y as in Definition 1, the sensitivity is given in terms of the diseased population. Let denote $\pi_i^k$ the conditional probability that the discrete random variable $X^k$ takes a fixed value $x_i$ given the diseased population, i.e. $\pi_i^k = P[X^k = x_i \mid Y=1]$. Hence according to (1) the sensitivity will be equal to,

$$Se(X^k) = \sum_{i=c}^{k} \pi_i^\kappa \quad (2)$$

**Proposition 1:** For any $k \in N$, the sensitivity, $Se(X^k)$, is a non-decreasing function as a function of k.

**Proof:** It needs to be established that for any $k \in N$

$$Se(X^k) \leq Se(X^{k+1}), \quad (3)$$

and according to (2) it is sufficient to show that

$$\sum_{i=c}^{k} \pi_i^\kappa \leq \sum_{i=c'}^{k+1} \pi_i^{\kappa+1} \quad (4)$$

The equality in (3) holds when c is chosen to be the median value for any $k \in N$. Consider a (k+1)-partition, then

$$\pi_i^{k+1} = \pi_i^k - a_i, \quad (5)$$

for some constant $a_i$, $i=1,...,k$. Furthermore one may argue without loss of generality *that* $c \leq c'$.

*Case 1: c = c'. According to (5),*

$$\sum_{i=c}^{k}\pi_i^k = \sum_{i=c}^{k}\left(\pi_i^{k+1} + a_i\right) \leq \sum_{i=c}^{k}\pi_i^{k+1} + \sum_{i=1}^{k}a_i = \sum_{i=c}^{k}\pi_i^{k+1} + \pi_{\kappa+1}^{\kappa+1} = \sum_{i=c}^{k+1}\pi_i^{k+1}$$

*And thus (4) has been established.*

*Case 2: c<c'.*

**Assumption 1:** *There exists* $b_i \in \mathbf{R}^+$ *such that* $\sum_{i=c}^{c'-1}\pi_i^k \leq \sum_{i=1}^{c'-1}b_i$

*Hence from (5) and Assumption 1, we have that*

$$\sum_{i=c}^{k}\pi_i^k = \sum_{i=c}^{c'-1}\pi_i^k + \sum_{i=c'}^{k}\pi_i^k \leq \sum_{i=1}^{c'-1}b_i + \sum_{i=c'}^{k}\left(\pi_i^{k+1} + a_i\right) \quad (6),$$

*Furthermore, from (5) and (6) and considering* $a_i = b_i$ *for any i=1,...,k, one may observe that*

$$\sum_{i=c'}^{k}\pi_i^{k+1} + \sum_{i=1}^{k}a_i = \sum_{i=c'}^{k}\pi_i^{k+1} + \pi_{k+1}^{k+1} = \sum_{i=c'}^{k-1}\pi_i^{k+1} \quad (7)$$

*Hence (6) and (7) complete the statement.* ∎

**Remark 2:** *Assumption 1 is required as a control property for the probabilities which are excluded from the (k+1)-partition, but included in the k-partition.*

## 3. Discussion

The aim of this work was to investigate whether the number of classes of a scale affects its diagnostic accuracy. It was proved that the sensitivity of a scale, which is a measure of diagnostic accuracy that shows how correctly the scale classifies the true diseased in medical research, is a non-decreasing function of the number of scale's classes.

Health measurement scales, and particularly the composite ones, are useful instruments in biomedical research, since they can quantify characteristics of the individuals that are difficult to measure, like behaviors, attitudes and beliefs (i.e., psychological symptoms like depression, severity of a disease, health-related quality of life, dietary habits etc). Despite the importance of these tools no specific methodology has been proposed for their development. One of the major unresolved issues is the choice of the number of classes a scale should have. Until now several scales have been developed using different methodologies, in order to measure the same characteristic, with a variety of scoring systems, starting from a binary coding to wider range. But, in all cases the choice of the number of classes was arbitrary.

In a recent publication by Kourlaba and Panagiotakos based on four simulation studies it was revealed that sensitivity of a scale is a increasing function of scale classes (Kourlaba and Panagiotakos, 2009b). In particular, a continuous scale was initially developed and afterwards thirteen other scales using 800-tiles, 500-tiles, 200-tiles, 100-tiles, 50-tiles, 15-tiles, percentiles, 8-tiles, 6-tiles, quintiles, quartiles, tertiles and median of the components of the initial scale. All scales were tested against a binary response outcome. Results based on 1,000 simulated data sets revealed that the maximum sensitivity of the scale was obtained when we use the maximum number of classes. However, simulations studies, although they can give an impression about the true relationships are always prone to bias. Thus, in this work, it was established that the sensitivity of a scale is a non-decreasing function of the number of classes used (or in other words the number of the values of the discrete variable) under a technical assumption (i.e., Assumption 1). This assumption provides a control on the mass of points which are considered in the $k$-partition, but excluded in the $(k+1)$-partition. It is open if this partition could actually improve the predicting ability of a simulated model and of an empirical paradigm.

At this point it should be noted that scales using small number of classes may be more comprehensible and easier applied in daily clinical practice; however, according to the simulated and proved results these scales result in low diagnostic accuracy. The use of small-range scales could also be an explanation for the weakness of some studies to detect a significant association between a health measurement scale and a disease outcome, where the association between what the scale aims to quantify and the evaluated outcome has long been understood (McCullough ML, 2000, Harnack, 2002). In the previously mentioned studies, small-scale indices which had been created using components with only two classes had been used.

## 4. References

1. Kant, AK. (1996). Indexes of overall diet quality: a review. Journal American Dietetic Association 96:785-91.

2. Kourlaba, G, Panagiotakos, DB. (2009). Dietary quality indices and human health: a review. Maturitas 62:1-8.

3. Kourlaba, G, Panagiotakos, DB. (2009) The diagnostic accuracy of a composite index increases as the number of partitions of the components increases and when specific weights are assigned to each component. Journal of Applied Statistics 37: 537-554.

4. McCullough ML, Feskanich D, Stampfer MJ, Rosner BA, Hu FB, Hunter DJ, Variyam JN, Colditz GA, Willett WC. (2000) Adherence to the Dietary Guidelines for Americans and risk of major chronic disease in women. American Journal Clinical Nutrition 72:1214-1222.

5. Harnack L, Nicodemus K, Jacobs DR, Jr., Folsom AR. (2002) An evaluation of the Dietary Guidelines for Americans in relation to cancer occurrence. American Journal Clinical Nutrition 76, 889-896.